# Application of Large Language Models for Container Throughput Forecasting: Incorporating Contextual Information in Port Logistics

Minseop Kim, Jaeeun Kwon, Hanbyeol Park, Kikun Park, Taekhyun Park, Hyerim Bae, *Member, IEEE*

*Abstract*—Recent advancements in generative artificial intelligence (AI) have demonstrated its substantial potential in various fields. However, its application in port logistics remains underexplored. Ports are complex operational environments where diverse types of contextual information coexist, making them a promising domain for the implementation of generative AI and highlighting the urgency of related research. In this study, we applied a large language model (LLM)—a leading generative AI technique—to forecast container throughput, which is a critical challenge in port logistics. To this end, we adopted a state-of-the-art LLM approach and proposed a novel prompt structure designed to incorporate the contextual characteristics of port operations. Extensive experiments confirm the superiority of our method, showing that the proposed approach outperforms competitive benchmark models. Furthermore, additional experiments revealed that LLMs can effectively learn and utilize multiple layers of contextual information for inference in port logistics. Based on these findings, we explore the key constraints affecting LLM adoption in this domain and outline future research directions aimed at addressing them. Accordingly, we offer both technical and practical insights to support the effective deployment of generative AI in port logistics.

## I. INTRODUCTION

THE recent rapid advancement in generative artificial intelligence (AI) extends it beyond natural language processing and demonstrates its substantial potential across a wide range of domain-specific applications. Owing to their broad linguistic knowledge and advanced reasoning capabilities, large language models (LLMs) have garnered considerable attention from both academia and industry [1], [2] [3], [4]. This growing interest stems largely from LLMs' general cognitive abilities, which enable them to integrate and interpret complex information via human-level language understanding and reasoning—capabilities that conventional AI models have struggled to achieve [5], [6]. Owing to their adaptability, LLMs can function as general-purpose decision-support tools across diverse fields, rather than being confined solely to language-related tasks.

Ports are inherently complex environments characterized by dynamic interactions among vessels, cargo, equipment, transportation systems, and human operators. In such settings, AI technologies have been employed to increase operational efficiency and support decision-making [7], [8], [9]. Owing to the powerful reasoning capabilities of LLMs, interest in their application in port logistics has recently increased. A few studies have highlighted the significance of these technologies in this context and proposed relevant research priorities [9], [10], [11]. Further, several studies in the transportation domain have conducted LLM-based demand-forecasting tasks and demonstrated their effectiveness [12]. This study shows how unstructured textual and semantic knowledge can be leveraged to improve forecasting performance. However, the number of researchers focusing on AI applications in port logistics remains limited [8], and research specifically addressing the integration of LLMs into this domain is still in its infancy. Despite these challenges, the transformative potential of generative AI offers novel solutions that were previously unattainable with conventional AI methods [9]. Consequently, demonstrating real-world applications that combine the scientific foundations of LLMs with the operational complexity of port logistics constitutes a timely and essential research direction. Such efforts can play a crucial role in bridging the gap between technological innovations and domain-specific challenges.

From this perspective, focusing on container throughput (CT)

This research was supported by Korea Institute of Marine Science & Technology Promotion (KIMST) funded by the Ministry of Oceans and Fisheries (RS-2022-KS221657) and the National Research Foundation of Korea (NRF) grant funded by the Korea government (MSIT)(No.RS-2023-00218913). *(Corresponding author: Hyerim Bae).* Minseop Kim, Jaeeun Kwon, Hanbyeol Park are with the Major in Industrial Data Science & Engineering, Department of Industrial Engineering, Pusan National University (e-mail: rlaals7349@pusan.ac.kr, kjy3865@pusan.ac.kr, pb104@pusan.ac.kr). Kikun Park is with Safe and Clean Supply Chain (SCSC) Research Center, Pusan National University (e-mail: kikunpark@pusan.ac.kr).

Hyerim Bae, Taekhyun Park are with the Graduate School of Data Science, Pusan National University (e-mail: hrbae@pusan.ac.kr, pthpark1@gmail.com).

forecasting, which is an ongoing research topic in the port-logistics domain, is essential. CT refers to the number of container-handling activities at a port, and numerous studies have attempted to forecast future CT by periodically aggregating such data and applying AI-based time-series forecasting (TSF) methods. These forecasts have been widely used to support key decision-making processes, such as resource allocation, supply chain planning, and hinterland logistics management, helping ports better manage operational uncertainties [14], [16], [17], [18], [19]. Recent studies have increased forecasting granularity to high-resolution levels, such as daily forecasts, and have begun using AI models that incorporate port operational characteristics to enhance their predictive performance. In this context, recent research has attempted to integrate various types of contextual information, such as berth schedules, time-of-day indicators, and environmental data, into forecast models. For example, [16] reported that contextual data significantly affect CT-forecasting accuracy. However, incorporating such contextual variables into conventional AI-based TSF models often introduces structural limitations. Specifically, this often requires converting multimodal data generated during port operations into rigid time-series structures. According to [21] and [22], this process can result in the loss of essential contextual semantics when the multimodal input is translated into continuous numerical values that conventional TSF models can interpret. Furthermore, existing AI-based TSF models often rely solely on historical numerical data and lack an inherent understanding of port logistics as a domain, thereby limiting their ability to produce context-aware forecasts. By contrast, LLMs have demonstrated the capacity to integrate and reason with multimodal information, supported by their advanced language-processing capabilities. This strength has been increasingly validated across multiple domains [6], [22], [23], demonstrating its potential to address complex and domain-specific forecasting challenges such as those encountered in port logistics.

Building on this research background, this study aims to explore a new area at the intersection of CT forecasting and LLMs by providing empirical evidence and practical implications. To this end, we propose an approach that employs an LLM to integrate diverse types of contextual information generated in port operations and forecast CT accordingly. We conducted extensive experiments using operational data collected from a real container terminal and analyzed both the predictive effectiveness of the LLM-based TSF approach and the inference mechanisms of the LLM through detailed post-hoc analysis. This study seeks to address the following three key research questions.

- **Research Question 1:** Are LLMs capable of achieving competitive forecast performance in forecasting CT?
- **Research Question 2:** Are LLMs capable of effectively learning and applying contextual information from port logistics?
- **Research Question 3:** Are LLMs practical tools for real-world port-logistics applications?

This study makes the following primary contributions. First, by incorporating port-logistics-specific knowledge into an LLM, we achieved CT forecasting that reflects diverse operational contexts and unstructured information. This represents a novel approach that simultaneously enhances forecasting performance and interpretability by leveraging contextual information that is typically difficult to capture in conventional AI-based TSF models. Second, we demonstrate the superiority of the proposed method through comparative experiments involving various state-of-the-art benchmarks and LLM-based models using real-world operational data. We further explored how the LLM internalizes port-logistics knowledge through post-hoc interpretability experiments. Third, we offer important implications for related domains by discussing the constraints encountered in applying LLMs to port logistics and proposing future research directions to address them.

This study holds substantial academic value as the first empirical study to apply LLMs to CT forecasting using a structured port-context prompt. Section 2 presents a detailed review of prior research on CT-forecasting and LLM-TSF methodologies. Section 3 introduces the LLM-TSF framework and outlines the design of prompts that embed domain-specific knowledge. Section 4 validates the proposed approach through comparative experiments against benchmark models and evaluates whether the LLM effectively learns port-logistics knowledge via post-hoc analysis. Section 5 discusses the limitations of the findings and outlines directions for future research. Section 6 concludes the study by revisiting the research questions and discussing the key findings and their broader implications.

## II. LITERATURE REVIEW

### A. CT Forecasting

CT is defined as the total number of containers handled by a port, and research in this field focuses on forecasting the volume of containers that a port will process in the future. Due to the high volatility inherent in CT, researchers have increasingly adopted or developed TSF models based on machine learning and deep learning. Early studies began applying AI techniques to forecast future CT based on historical CT, aiming to capture the complex nonlinear dynamics in CT patterns. For example, [17] and [30] utilized long short-term memory (LSTM), a widely used AI technique in TSF, to forecast CT. Various other AI-based TSF methodologies have also been explored, including Convolutional Neural Networks (CNN)-LSTM [29] and temporal convolutional networks [38].

Subsequent studies developed methods aimed at capturing the complex nonlinear characteristics of CT. Reference [31] proposed a novel approach that combines genetic algorithms and simulated annealing (SA) to optimize the architecture of artificial neural networks (ANNs) for CT forecasting. Reference [13] demonstrated state-of-the-art performance by combining decomposition techniques with machine learning methods. Reference [19] introduced a method that integrates data characteristics analysis (DCA) and least squares support vector regression (LSSVR), [37] proposed a methodology combining discrete wavelet transform (DWT) and LSTM to enhance CT forecasting performance. This line of research reflects a progression from the initial application of AI models toward the development of methods capable of better representing the complex nonlinearity of CT [25], [26], [28], [33], [35], [39].

Another stream of research has sought to incorporate various external factors into forecasting models. Reference [36] extracted economic, environmental, and social factors and applied LSTM within a multivariate time-series forecasting framework to account for their influence on CT. Reference [32] hypothesized that the Industrial Confidence Indicator for the Euro Area (ICIEA) serves as a leading indicator of CT and integrated it into the forecasting model. Reference [60] forecasted CT for Chinese ports based on nine indicators, including regional gross domestic product, whereas [61] selected twelve socio-economic factors and incorporated them into forecasting models. Several other studies have similarly attempted to reflect external influencing factors [27], [33], [34], [36]. These efforts represent a parallel research direction aimed at integrating external determinants of CT to improve forecasting performance.

Recently, [16] emphasized the need for CT forecasting to support real-time decision-making in daily port operations. They employed XGBoost and ARIMA models incorporating time-related information, vessel workload, and day-of-week variables to forecast CT data for the following day at a container terminal. They argued that such forecasting could assist practitioners in making optimal decisions during daily operations and help reduce operational costs. A key contribution of this study is its shift from prior research focused on CT at a macroscopic scale, such as monthly or yearly aggregations, toward daily CT forecasting that can support day-to-day operational decisions. This transition raises the need to effectively integrate diverse high-resolution factors arising from daily port operations. In essence, it highlights the necessity of forecasting models capable of comprehensively understanding and incorporating the operational context of port logistics. However, conventional machine learning and AI-based approaches, such as those adopted in [16], require various operational data to be preprocessed and converted into numerical features before being fed into the model. This dependence on processed inputs restricts the model's ability to understand the underlying operational context of port logistics, as it merely computes numerical relationships without interpreting the semantic implications of the data.

Building on prior work, we anticipate that applying LLMs to CT forecasting can enhance performance in two key ways. First, LLMs can holistically interpret diverse information related to port operations—vessel schedules and weather reports—expressed in natural language, thereby enabling the model to comprehend operational contexts and forecast CT. Second, because this contextual information can be supplied as text prompts without extensive preprocessing, the approach substantially simplifies the large-scale data processing required by conventional TSF models. Consequently, this study extends existing research by introducing an LLM-based method that directly incorporates port-logistics knowledge to forecast CT.

## B. LLM-TSF

TSF has long been a central research topic across various domains, including port logistics, climate modeling, traffic management, healthcare monitoring, and financial analytics [8] [40]. To meet this demand, researchers have developed models using both statistical techniques and deep learning approaches. In recent years, particular attention has been given to methodologies incorporating the attention mechanism, which demonstrates excellent performance in translation tasks [43], [44], [45], [46]. These studies commonly aim to forecast future points or intervals based on historical time series (TS) data, uncover complex temporal dependencies, and effectively model interactions among multiple variables.

Recently, in the field of TSF, LLM-TSF approaches—those that leverage LLM to forecast TS data—have shown considerable potential, as demonstrated by numerous studies [1], [2], [3], [40], [47], [48], [49]. These studies have emerged owing to the limitations of conventional AI-based methodologies, which often rely on only segments of historical numerical data, thereby constraining their ability to perform intrinsic reasoning in TS [20], [15]. Here, intrinsic reasoning refers to the use of contextual understanding—background knowledge and domain-specific insights—to inform forecasting [6]. LLMs excel at reasoning and pattern recognition within complex, structured domains. Therefore, interpreting text-based domain knowledge relevant to TS and leveraging it for forecasting represents a promising research direction.

Therefore, in this section, we systematically review and analyze recent developments in the field of LLM-TSF from two key perspectives: (1) prompt design, (2) cross-modality.

### 1) Prompt Design

Prompt design is an approach that designs prompts to improve TSF performance by providing explicit instructions or contextual information to LLMs [55] [56]. This approach was proposed to address the limitations of conventional TSF methods that rely on unimodal numerical time-series data [21]. Several recent studies have explored enhancing forecasting performance by embedding multiple layers of relevant contextual information into LLMs through prompt engineering [22] [40] [50] [53] [54]. Reference [21] emphasized the value of text-based contextual inputs by scraping 10 articles and using ChatGPT 3.5 to generate summaries, which were then used as prompts to enrich the model's forecasting context. This finding demonstrates that integrating text-based knowledge closely tied to forecasting tasks can significantly improve model performance. Similarly, [15] introduced a prompt-as-prefix strategy, which incorporates domain knowledge, task-specific details, and statistical descriptions of input data into prompts. Their findings suggest that forecasting in specialized domains can be improved by injecting such domain-specific contextual information.

In addition, studies across various domains have sought to enhance the forecasting performance of LLMs by explicitly incorporating domain knowledge. Reference [3] improved the forecasting performance in the healthcare domain by feeding clinical reports into LLMs alongside TS data to perform diagnostic tasks using ECG signals. Similarly, [2] demonstrated the reasoning capabilities of an LLM by first pretraining it by converting large-scale clinical data into textual data and utilizing it for various downstream tasks. In the financial domain, [47] and [59] effectively utilized contextual information related to stocks by inputting historical economic and financial news into an LLM for TS forecasting.

These studies reflect two converging research directions: one focuses on developing new methodologies to help LLMs understand and leverage prompts more effectively; the other explores prompt engineering tailored to solve domain-specific forecasting problems. This study aligns with the latter direction and introduces a novel prompt design to address a practical task in port logistics.

### 2) Cross-Modality with Contextual Information

Prompt design focuses on determining what information to provide, whereas another stream of research investigates how to effectively align different data modalities from a cross-modality perspective. Reference [23] introduced a language-TS Transformer to enable cross-modal learning between TS data and human-generated domain instructions. References [15] and [53] proposed an approach that integrates tokenized prompts and TS using a multi-head self-attention (MHSA) mechanism, achieving state-of-the-art forecasting performance. Reference [5] proposed a technique to reduce word embedding using principal component analysis (PCA) and introduced contrastive loss to align the resulting representations with TS data. Reference [15] extracted text prototypes—a curated subset of natural language—from word embeddings and developed a reprogramming method to align them with TS, thereby achieving cross-modality and improving the forecasting performance. Reference [6] utilized semantic anchors derived from word embeddings to learn a shared latent space with TS, further advancing cross-modality integration. Collectively, these studies demonstrate a growing interest in developing methodologies that bridge the numerical time series data and semantic representation.

## III. METHODOLOGY

### A. Methodology Selection

In this section, based on the analysis of prior studies presented in Section 2, we adopt the TimeLLM methodology proposed by [15] guided by three key considerations: 1) flexible prompt structure, 2) computational efficiency, and 3) interpretability. First, the prompt-as-prefix approach is designed with a flexible structure capable of incorporating diverse contextual information related to port operations. Second, by freezing the pretrained parameters of the LLM and training only a limited number of added layers, the model enables rigorous experimentation and performance validation using relatively few computational resources. Third, a key advantage of TimeLLM lies in its interpretability, which is supported through text prototype construction, reprogramming, and layerwise visualization. This makes it easy to evaluate whether the LLM has learned port-logistics contextual information. In this study, we adopt the Qwen 3B model [57] due to its computational efficiency and accessibility. We integrate the proposed port logistics knowledge prompt (PK-prompt) into the TimeLLM architecture and refer to the resulting framework as PK-TimeLLM. Fig. 1 provides a high-level overview of the proposed methodology.

### B. PK-TimeLLM

### 1) Text Prototyping

To present words, LLMs utilize a word-embedding matrix $E$, as defined in Eq. (1). Here, $V$ denotes the vocabulary size, and $D$ represents the dimension of the embedding vector corresponding to each word. The complete embedding space is thus represented as $\mathbb{R}^{V \times D}$, where each of the $V$ words are embedded as a vector in an $\mathbb{R}^D$ dimension.

$$E \in \mathbb{R}^{V \times D} \qquad (1)$$

The text prototype refers to a small subset of semantically meaningful words extracted from $E$, which are specifically constructed to emphasize terms critical for forecasting during training. The full matrix $E$ is not used directly because it contains a large number of tokens that are irrelevant to forecasting, which can increase the computational cost and introduce noise, ultimately degrading the forecasting performance. To address this, a weight matrix $W^E \in \mathbb{R}^{V' \times V}$ is applied, as shown in Eq. (2), to embed $E$ into a text prototype. This enables the model to prioritize forecasting-relevant terms, thereby enhancing both the computational efficiency and inference accuracy. The resulting text prototype is denoted by $E`$, as defined in Eq. (3), where the vocabulary is reduced to a smaller size, $V'$.

$$W^E \in \mathbb{R}^{V' \times V} \qquad (2)$$

$$E` = W^E E \in \mathbb{R}^{V' \times D} \qquad (3)$$

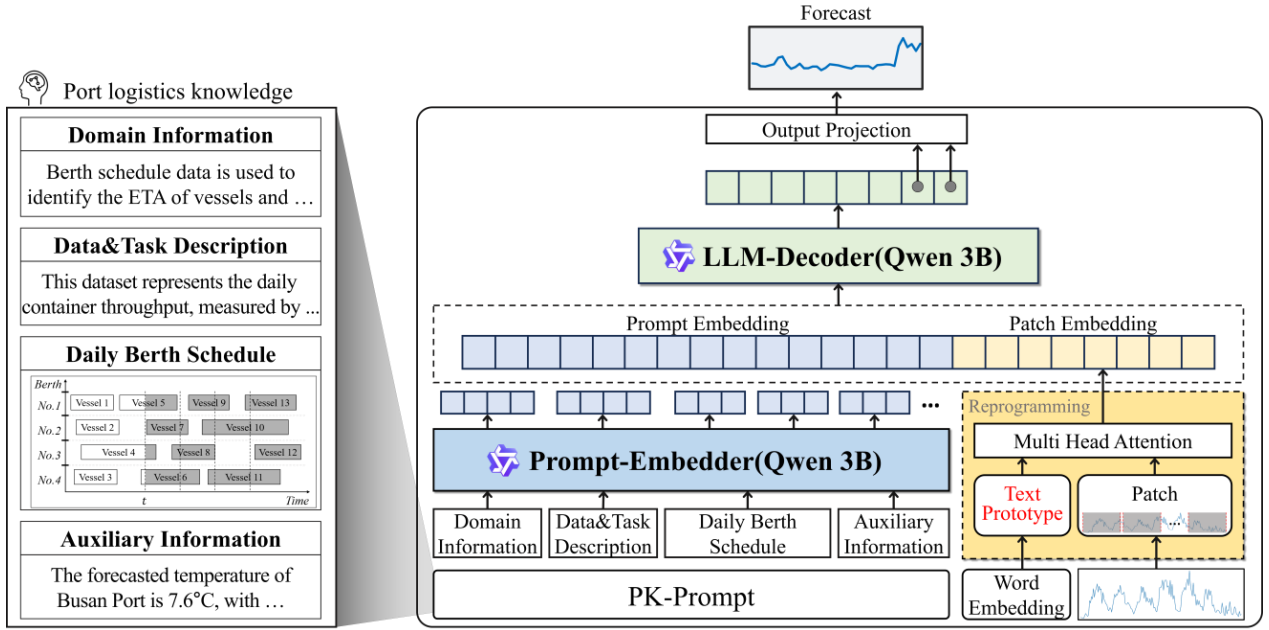

**Fig. 1.** Framework of PK-TimeLLM

## 2) Reprogramming

The reprogramming step aligns the historical CT sequence with the text prototype.

$$X_t = (x_1, x_2, \ldots, x_t) \in \mathbb{R}^T \quad (4)$$

Let $X_t$ denote the historical CT sequence defined in Eq. (4), where $x_t$ is the daily container throughput observed at time step $t$.

$$X_P \in \mathbb{R}^{P \times L_P} \quad (5)$$

$$\hat{X}_P = W^Q X_P \in \mathbb{R}^{P \times d_m} \quad (6)$$

$$K = E^{\backprime} W^K \in \mathbb{R}^{V^{\backprime} \times d_m} \quad (7)$$

$$V = E^{\backprime} W^V \in \mathbb{R}^{V^{\backprime} \times d_m} \quad (8)$$

$X_t$ is partitioned into consecutive patches, denoted as $X_P$, as shown in Eq. (5). Let $L_P$ represent the length of each patch and $S$ the stride size used when generating patches, the total number of patches $P$ is calculated as $\frac{(T - L_P)}{S} + 2$. Each patch is embedded into a $d_m$-dimensional vector space using a learnable weight matrix $W^Q$, resulting in $\hat{X}_P$, as formulated in Eq. (6). Similarly, to embed the text prototype into the same dimensional space as $\hat{X}_P$, weight matrices $W^K, W^V \in \mathbb{R}^{D \times d_m}$ are introduced, as described in Eqs (7) and (8), respectively.

$$Z = Softmax\left(\frac{\hat{X}_P K^\top}{\sqrt{d_m}}\right) V \in \mathbb{R}^{P \times d_m} \quad (9)$$

$\hat{X}_P$, $K$, and $V$ execute the MHSA mechanism, as described in Eq. (9), which can be interpreted as a question–answer relationship. Specifically, $\hat{X}_P$ derived from $X_t$ serves as the question, whereas $K$, generated from $E^{\backprime}$, serves as the answer. The MHSA mechanism calculates the similarity between the question and the answer, integrates this with $V$, and outputs a result denoted as $Z \in \mathbb{R}^{P \times d_m}$. This step effectively combines the linguistically informed text prototype with the $X_t$ to align them in a shared representation space.

$$W^O \in \mathbb{R}^{d_m \times D} \quad (10)$$

$$Patch\ Embedding = W^O Z \in \mathbb{R}^{P \times D} \quad (11)$$

Finally, $Z$ is embedded into a $D$-dimensional token that the LLM can interpret. This step is commonly referred to as tokenization, as it converts the output into the embedding space used by the LLM to represent words as vectors in $\mathbb{R}^D$. To achieve this, the weight matrix $W^O \in \mathbb{R}^{d_m \times D}$ defined in Eq. (10) is applied to produce $P$ tokens, which constitute the patch embedding, as shown in Eq. (11).

## 3) Port-Logistics Knowledge Prompt (PK Prompt)

In this section, we propose a PK prompt designed to integrate contextual information relevant to CT forecasting. The prompt was developed based on prior research on CT forecasting and refined through consultations with domain experts working at container terminals. An example of a PK prompt is presented in Table I.

The Data Description component provides meta-information about the dataset, including the data collection period, geographical location, and aggregation method. The Task Description component outlines the specific forecasting objectives and describes the rationale for the strong association between CT and berth schedules. It also explains typical patterns observed in port operations, such as the tendency for CT to decrease on weekends or public holidays. By embedding this contextual information, the PK prompt enables the LLM to better understand the relationships between background factors and the forecasting target.

TABLE I
EXAMPLE OF PK PROMPT

| Category | Contents |
|---|---|
| *Data Description* | This dataset captures the daily CT, measured by gate-in and gate-out activities at a container terminal in Busan Port, spanning the period from January 2022 to December 2022. |
| *Task Description* | The input data consist of historical CT. Your task is to forecast the next $< H >$ steps based on the previous $< T >$ steps, in collaboration with the provided prompting information |
| *Domain Information* | Berth schedule data are used to identify the estimated time of arrival of vessels and the anticipated CT. CT generally increases with the number of scheduled vessel arrivals, as a higher number of vessels typically requires more operational activity. Throughput often declines on weekends, and public holidays owing to reduced road truck activity and may also decrease on days with heavy rainfall because of weather-related disruptions. |
| *Berth Schedule* | FOR each step $h$ in $< H >$ DO # (where $H$ is the forecasting horizon):<br>"Forecasting step $< h >$($date_h$, $week_h$)<br>$< date_h >$ is expected to have a [$low$\|$average$\|$high$\|$very high$] operational volume, as $< n_h >$ vessel(s) are scheduled to arrive, with an estimated loading/unloading volume of $< I_h + E_h >$ TEUs."<br>END<br>"After next $< H >$ days, Estimated vessels to arrival are $< \sum_{h=1}^{H}(n_h) >$, Volumes are $< \sum_{h=1}^{H}(I_h + E_h) >$" |
| *Auxiliary Information* | FOR each step $h$ in $< H >$ DO # (where $H$ is the forecasting horizon):<br>"$< date_h >$ is a [$working\ day$\|$weekend$\|$holiday$], and the name of *the* holiday is $< holiday\ name >$.<br>The forecasted temperature of Busan port is $< temperature_h >$°C, with a precipitation of $< precipitation_h >$ mm and an expected wind speed of $< windspeed_h >$ m/s."<br>END |

The Berth schedule component includes details of scheduled vessels and their associated workloads, which serve as external factors with a strong influence on CT. The significance of this information has been emphasized by both domain experts and prior research [16]. Arrival schedules and expected handling volumes of vessels represent the anticipated operational workload at the port, which directly affects CT. In practice, many container terminals use this information to formulate operational plans for assigning equipment and personnel. The Berth Schedule component generates prompts from berth schedule data, indicating the number of incoming vessels $n_h$ and the estimated import/export container volumes $I_h$ and $E_h$ at each forecasting horizon step $h$. This process was repeated across the entire forecasting horizon. The berth plan information included in the PK prompt is fully available at the time of forecasting because all vessels are required to pre-notify the container terminal of this information before arrival.

Finally, the Auxiliary Information component incorporates additional variables, such as wind speed, temperature, precipitation, and day of the week. According to domain experts, CT tends to decrease on weekends and public holidays, or under heavy precipitation conditions. In some cases, port operations may be suspended owing to safety concerns during high-wind conditions, resulting in a sharp decrease in CT. These environmental factors are irregular and difficult to interpret using conventional TSF models. Therefore, they were embedded into the PK prompt to allow the LLM to account for them more effectively. All information other than the Berth schedule is likewise structured to be available at the forecast time and is neither delayed nor missing. The proposed PK prompt is embedded through the Prompt embedder and has the dimension of $\mathbb{R}^{N_T \times D}$, as shown in Eq. (12).

$$Prompt\ Embedding \in \mathbb{R}^{N_T \times D} \quad (12)$$

$$R \in \mathbb{R}^{(N_T + P) \times D} \quad (13)$$

$$\hat{y} = f(R): \mathbb{R}^{(N_T + P) \times D} \rightarrow \mathbb{R}^H \quad (14)$$

Here, $N_T$ denotes the total number of tokens generated from the PK prompt, which is determined by the number of words used. $R$, defined in Eq. (13), represents the combined embedding of the prompt and the patch, which is then passed to the LLM-Decoder, as shown in Eq. (14). The LLM-Decoder receives $R$ as input and generates forecast values over the forecasting horizon $H$.

## IV. EXPERIMENTS

### A. Experimental Setting

For these experiments, we selected Busan port and collected operational data spanning January 1, 2022, to December 31, 2023. The dataset was split by year to create two experimental datasets. Daily CT, denoted as $CT_t$ as shown in Eq. (15), represents the total number of containers either inbound to or outbound from the container terminal during day $t$.

In Eq. (16), $T$ denotes the input length, and $H$ denotes the

$$CT_t = \sum_{c_t \in [t, t+1)} 1 \quad (15)$$

$$\widehat{CT}_{t+1:t+H} = f_\theta(CT_{t-T+1:t}) \quad (16)$$

forecasting horizon. We adopt this multi-step forecasting formulation, which map the past $T$ observations to CT values over the next $H$ steps, extending beyond the single-step forecasting ($H$=1) in [16] and posing a more challenging task that requires the model to capture longer-range temporal dependencies.

A total of 18 experiments were conducted by varying the $T$ (14, 21, and 28 days) and $H$ (1, 7, and 14 days) to assess the model's generalization ability. We compared the proposed method with nine state-of-the-art benchmark models representing a range of forecasting architectures, including statistical [66], recurrent [41], convolutional [42], [65], multilayer perceptron [63], linear [62], transformer [64], and LLM-based models [51], [52]. For the benchmark models, the hyperparameters were selected through a grid search, and the model with the best validation performance was retained. All datasets were strictly divided along the temporal axis to prevent information leakage and normalized using standard scaling. The dataset was divided into training, validation, and test sets in a 50:30:20 ratio. Early stopping was employed, terminating training if the validation loss failed to improve over ten consecutive iterations. The training process used the mean squared error (MSE) as the loss function, which was calculated based on the difference between the forecasted and actual CT. Both MSE and mean absolute error (MAE), which are widely used in CT forecasting, were employed to evaluate the model's performance.

$$MSE = \frac{\sum(\hat{y} - y)^2}{n} \quad (16)$$

$$MAE = \frac{\sum|\hat{y} - y|}{n} \quad (17)$$

Details of the hyperparameters and model configurations are provided in Appendix 1.

### B. Experimental Results

#### 1) Performance Comparison with Benchmark Models

Table II summarizes the results of 18 experiments conducted using combinations of input length $T$ and forecasting horizon $H$ across two datasets. Model performance was evaluated using MSE, with the best-performing model in each experiment highlighted in bold and the second-best indicated with underlining.

Initially, the seasonal autoregressive integrated moving average (SARIMA) model, which relies on statistical techniques, exhibited low accuracy under various experimental conditions. This result stems from SARIMA's dependence on historical data patterns, such as autocorrelation and moving averages, which are insufficient for capturing the nonlinear characteristics of CT. Among the recurrent unit-based models, LSTM achieved relatively low errors when the forecasting horizon was set to $H = 1$; however, its performance declined significantly as the input length increased, highlighting its limitations in long-range forecasting.

LSTnet showed increasing errors with longer horizons in Dataset 1 but maintained a relatively stable performance in Dataset 2. This stability is attributed to its skip-connection mechanism, which enables the model to selectively focus on the most relevant segments of the input rather than learning the entire CT sequence uniformly. TSMixer, a forecasting model based on a multilayer perceptron (MLP) architecture, is designed to capture variable interactions through feature mixing.

TABLE II
EXPERIMENTAL RESULTS

| | Models | $T = 14$ | | | $T = 21$ | | | $T = 28$ | | |
| | | $H = 1$ | $H = 7$ | $H = 14$ | $H = 1$ | $H = 7$ | $H = 14$ | $H = 1$ | $H = 7$ | $H = 14$ |
|---|---|---|---|---|---|---|---|---|---|---|
| **Dataset 1** | SARIMA | 0.658 | 0.894 | 1.034 | 0.522 | 1.057 | 1.282 | 0.696 | 1.089 | 1.191 |
| | LSTM | 0.482 | 0.615 | 1.006 | 0.428 | 0.795 | 1.142 | 0.978 | 1.069 | 1.062 |
| | LSTnet | 0.492 | 0.787 | 0.941 | 0.504 | 0.923 | 1.011 | 0.694 | 0.882 | 0.883 |
| | DLinear | 0.486 | 0.754 | 0.894 | 0.544 | 0.817 | 0.996 | 0.736 | 0.678 | 0.803 |
| | PatchMixer | 0.468 | <u>0.708</u> | 0.872 | <u>0.419</u> | <u>0.690</u> | <u>0.826</u> | **0.377** | 0.708 | <u>0.624</u> |
| | PatchTST | <u>0.357</u> | 0.713 | 0.885 | 0.470 | 0.772 | 0.957 | 0.470 | 0.723 | 0.674 |
| | TSMixer | 0.638 | 0.764 | <u>0.815</u> | 0.566 | 0.804 | 0.893 | 0.542 | 0.781 | 0.773 |
| | AutoTimes | 0.632 | 0.880 | 1.022 | 0.651 | 0.803 | 1.087 | 0.661 | 0.818 | 0.968 |
| | aLLM4TS | 0.423 | 0.779 | 1.020 | 0.440 | 0.771 | 0.893 | 0.454 | <u>0.657</u> | 0.806 |
| | **PK-TimeLLM** | **0.331** | **0.588** | **0.655** | **0.347** | **0.563** | **0.584** | **0.406** | **0.491** | **0.542** |
| | Avg | 0.479 | 0.732 | 0.901 | 0.485 | 0.771 | 0.932 | 0.591 | 0.756 | 0.793 |
| **Dataset 2** | SARIMA | 0.897 | 0.767 | 0.789 | 0.349 | 0.365 | 0.516 | 0.390 | 0.276 | 0.309 |
| | LSTM | 0.299 | 0.212 | 0.167 | 0.436 | 0.219 | 0.210 | 1.176 | 0.718 | 1.193 |
| | LSTnet | 0.208 | 0.258 | 0.240 | 0.177 | 0.245 | 0.242 | 0.188 | 0.260 | 0.248 |
| | DLinear | 0.211 | 0.252 | 0.255 | 0.236 | 0.285 | 0.264 | 0.280 | 0.396 | 0.326 |
| | PatchMixer | 0.311 | 0.176 | 0.159 | 0.201 | 0.210 | <u>0.174</u> | 0.304 | 0.318 | 0.234 |
| | PatchTST | 0.181 | <u>0.170</u> | <u>0.156</u> | 0.192 | 0.189 | 0.178 | 0.205 | 0.320 | 0.236 |
| | TSMixer | 0.183 | 0.223 | 0.221 | 0.201 | 0.234 | 0.286 | 0.240 | 0.336 | 0.368 |
| | AutoTimes | 0.730 | 0.215 | 0.177 | 0.514 | 0.277 | 0.198 | 0.849 | <u>0.219</u> | <u>0.231</u> |
| | aLLM4TS | <u>0.165</u> | 0.180 | 0.479 | <u>0.174</u> | <u>0.188</u> | 0.487 | <u>0.199</u> | 0.462 | 0.511 |
| | **PK-TimeLLM** | **0.090** | **0.155** | **0.138** | **0.058** | **0.169** | **0.140** | **0.057** | **0.191** | **0.143** |
| | Avg | 0.328 | 0.261 | 0.278 | 0.254 | 0.238 | 0.270 | 0.389 | 0.350 | 0.380 |

However, despite its advanced design, it underperformed because it was trained solely on univariate CT data. TSMixer, a forecasting model based on a multilayer perceptron (MLP) architecture, is designed to capture variable interactions through feature mixing. By contrast, PatchMixer and PatchTST demonstrated competitive performance across both datasets. This result likely reflects the effectiveness of their patchwise training strategy, which segments CT into localized patterns and captures temporal dependencies more effectively.

LLM-based approaches, such as AutoTimes and aLLM4TS, demonstrated different performance trends across the two datasets. Whereas PatchMixer and PatchTST achieved superior forecasting accuracies on Dataset 1, AutoTimes and aLLM4TS performed more competitively on Dataset 2. However, PK-TimeLLM consistently achieved the best performance under most experimental conditions. In Dataset 1, it ranked highest in eight of the nine experiments, showing an average improvement of 24.26% over the second-best model. For Dataset 2, PK-TimeLLM outperformed all the other models across all the experiments, achieving an average improvement of 46.89% over the next-best performer.

Conventionally, in TSF, forecasting errors tend to increase as the forecasting horizon is extended. This trend arises because uncertainty accumulates when forecasts are generated sequentially using a limited amount of historical data. For example, in the first dataset under the conditions $T = 28$ and $H = 1$, PatchMixer achieved the best performance with an MSE of 0.377. However, when the forecasting horizon increased to 14, the MSE rose to 0.624, reflecting a performance decrease of approximately 40%. By contrast, PK-TimeLLM maintained a robust performance under the same conditions. Its MSE increased from 0.406 to 0.542, representing a smaller relative decrease of approximately 25%. A similar trend was observed for the second dataset. The performance of aLLM4TS, which was the second-best model under that setting, declined significantly, with its MSE rising from 0.199 to 0.511. Although PK-TimeLLM also experienced an MSE increase—from 0.057 to 0.143—its initial performance at $H = 1$ was already substantially better. Specifically, its MSE at the beginning of the forecasting horizon was 72% lower than that of aLLM4TS, confirming PK-TimeLLM's robustness across both short- and long-term horizons. Conventional TSF models rely solely on historical CT as input, making it difficult to incorporate dynamically changing contextual information. To illustrate this limitation, Fig. 2 presents the forecasting results under an experimental setting with an input length of T=28 and a forecasting horizon of H=14.

The figure shows a subset of results from two datasets, with comparisons made across AutoTimes, aLLM4TS, PatchMixer, PatchTST, and PK-TimeLLM—all of which demonstrated competitive performance in the experiments. Each data point represents a daily CT. Fig. 2(a) illustrates a case from Dataset-1 where benchmark models failed to accurately forecast future CT. This suggests that relying exclusively on historical CT does not capture sufficient patterns for longer-horizon forecasting. Fig. 2(b) depicts a scenario with two pronounced cycles of decline and subsequent surge, during which CT increased by nearly threefold compared to the previous day. In this case, PK-TimeLLM outperformed the benchmark models, successfully capturing these abrupt fluctuations. This improvement can be attributed to the incorporation of contextual information via PK prompts, which enabled PK-TimeLLM to more precisely forecast whether CT would rise or fall under varying conditions.

### 2) Multivariate Forecasting Incorporating Contextual Information

This section explores whether incorporating contextual information from the PK prompt can enhance the performance of conventional TSF models when used as additional input features. For this experiment, we generated variables from the PK prompt that could be transformed into a time-series format.

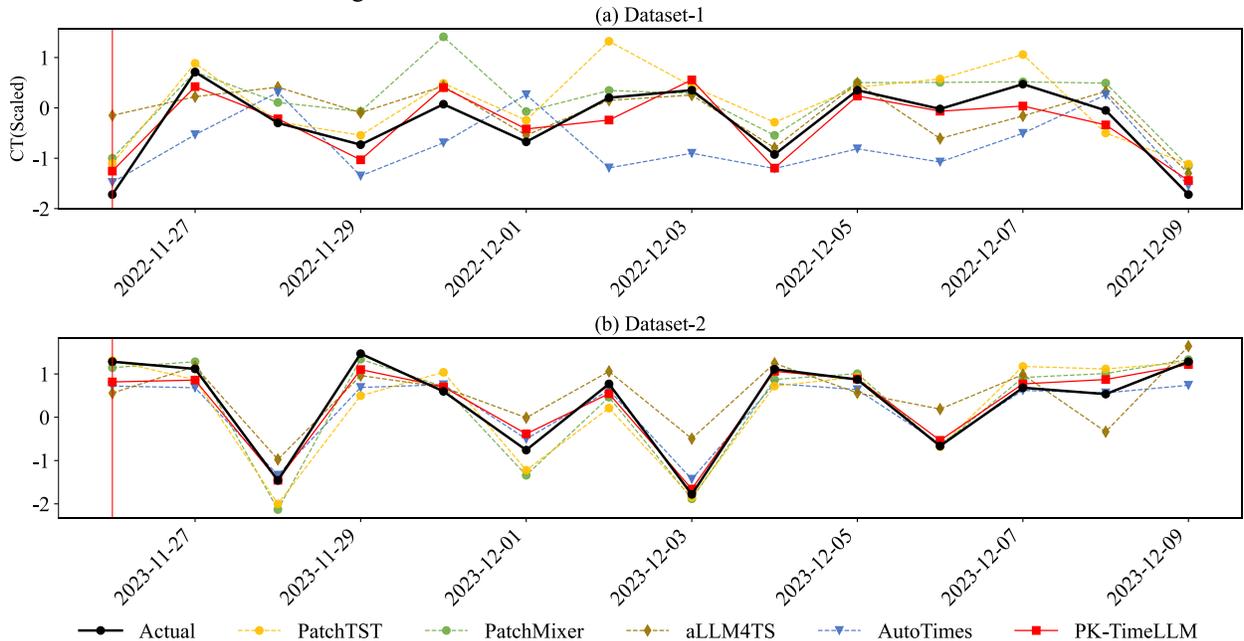

**Fig. 2.** Comparison of the forecasting results of each benchmark model (P = 14)

TABLE III
MULTIVARIATE FORECASTING RESULTS

| Models | | T = 14 | | | T = 21 | | | T = 28 | |
| | U | M | Imp | U | M | Imp | U | M | Imp |
|---|---|---|---|---|---|---|---|---|---|
| **Dataset 1** | | | | | | | | | |
| LSTM | 0.701 | 1.369 | -48.79% | 0.788 | 1.192 | -33.89% | 1.036 | 1.172 | -11.60% |
| LSTnet | 0.740 | 1.305 | -43.30% | 0.813 | 1.365 | -40.44% | 0.820 | 1.314 | -37.60% |
| DLinear | 0.711 | 0.695 | 2.30% | 0.786 | 0.739 | 6.36% | 0.739 | 0.604 | 22.35% |
| PatchMixer | 0.683 | 0.682 | 0.15% | <u>0.645</u> | 0.653 | -1.23% | <u>0.570</u> | <u>0.575</u> | -0.87% |
| PatchTST | <u>0.652</u> | <u>0.661</u> | -1.36% | 0.733 | <u>0.718</u> | 2.09% | 0.622 | 0.622 | 0.00% |
| TSMixer | 0.739 | 0.698 | 5.87% | 0.754 | 0.771 | -2.20% | 0.699 | 0.699 | 0.00% |
| AutoTimes | 0.845 | 0.960 | -11.98% | 0.847 | 0.905 | -6.41% | 0.816 | 0.814 | 0.25% |
| aLLM4TS | 0.741 | 0.960 | -22.81% | 0.701 | 0.902 | -22.28% | 0.639 | 0.752 | -15.03% |
| **PK-TimeLLM** | **0.525** | **0.525** | 0% | **0.498** | **0.498** | 0% | **0.480** | **0.480** | 0% |
| Avg | 0.727 | 0.916 | -20.71% | 0.758 | 0.906 | -16.26% | 0.743 | 0.819 | -9.33% |
| **Dataset 2** | | | | | | | | | |
| LSTM | 0.226 | 0.967 | -76.63% | 0.288 | 0.843 | -65.84% | 1.029 | 0.898 | 14.59% |
| LSTnet | 0.235 | 0.879 | -73.27% | 0.221 | 0.829 | -73.34% | 0.232 | 0.823 | -71.81% |
| DLinear | 0.239 | 0.224 | 6.70% | 0.262 | 0.260 | 0.77% | 0.334 | 0.306 | 9.15% |
| PatchMixer | 0.215 | 0.228 | -5.70% | 0.195 | <u>0.187</u> | 4.28% | 0.286 | 0.305 | -6.23% |
| PatchTST | <u>0.169</u> | <u>0.160</u> | 5.63% | <u>0.187</u> | 0.191 | -2.09% | <u>0.254</u> | <u>0.242</u> | 4.96% |
| TSMixer | 0.209 | 0.263 | -20.53% | 0.240 | 0.302 | -20.53% | 0.315 | 0.294 | 7.14% |
| AutoTimes | 0.374 | 1.364 | -72.58% | 0.330 | 1.219 | -72.93% | 0.433 | 1.399 | -69.05% |
| aLLM4TS | 0.275 | 1.565 | -82.43% | 0.283 | 1.530 | -81.50% | 0.391 | 1.449 | -73.02% |
| **PK-TimeLLM** | **0.128** | **0.128** | 0% | **0.122** | **0.122** | 0% | **0.130** | **0.130** | 0% |
| Avg | 0.243 | 0.706 | -65.63% | 0.251 | 0.670 | -62.58% | 0.409 | 0.715 | -42.72% |

These included the number of arriving vessels, handling volume, day of the week, holiday indicator, precipitation, wind speed, and temperature. The same aggregation method described in Section 3.B was applied to convert these variables into continuous numerical values. All benchmark models were retrained using these extended inputs. As PK-TimeLLM already integrates the same contextual data via the PK prompt, no new experiments were conducted for it. SARIMA was excluded from this experiment because it is a univariate model and cannot directly process multivariate inputs. Table III presents the experimental results, where "U" refers to univariate input and "M" refers to multivariate input. All outcomes are reported as the average MSE across the three forecasting horizons.

The results indicated that most models did not exhibit significant improvements in forecasting performance when additional variables were incorporated; these variables often acted as noise and led to performance degradation. DLinear was the only model that demonstrated consistent performance gains across all conditions, which may be attributed to its architecture that avoids transforming input data into higher-dimensional representations—thereby making it more robust to noise from multiple variables. Although PatchMixer and PatchTST maintained competitive performance, they did not outperform their univariate counterparts, suggesting that multivariate information did not yield meaningful gains. LSTnet and TSMixer, which are specifically designed to model interactions among variables, experienced a decline in performance, indicating that the added features may not have had strong interrelationships and instead contributed noise. This suggests that the models overfitted irrelevant information. Similarly, LLM-based models such as AutoTimes and aLLM4TS also exhibited performance deterioration, reinforcing the notion that the inclusion of these additional variables primarily introduced noise, rather than a beneficial context. Overall, the performance of the benchmark models declined across all the experimental settings, suggesting that they were unable to effectively capture the contextual information embedded in the additional variables. Although contextual data were incorporated into conventional TSF models, converting them into continuous sequences—a common practice—may have contributed to suboptimal performance. By contrast, PK-TimeLLM, which utilizes this contextual information through structured prompts, continued to deliver strong forecasting accuracy. This outcome highlights that supplying complex contextual information in prompt form is an effective strategy for enabling models to reflect rich operational nuances, thereby allowing LLM-based approaches to outperform conventional models.

### 3) Effectiveness of PK Prompt

In this section, we analyze whether incorporating contextual information from port logistics enhances forecasting performance. To this end, three scenarios were compared: (1) no prompt was provided, (2) a static prompt that excludes dynamic components (i.e., Berth Schedule and Auxiliary Information in Table I) was provided—this scenario omits dynamically changing contextual information from port operations and instead applies only varying statistical summaries of the historical CT, consistent with prior research [15]. The prompt used in this scenario is detailed in Appendix B. (3) The proposed PK prompt was applied. Table IV summarizes the forecasting performance across the three prompt configurations.

In 18 experiments, the second configuration did not consistently improve performance compared with the first configuration, which lacked prompts.

TABLE IV
PERFORMANCE COMPARISON ACROSS PROMPT CONFIGURATIONS

| | | w/o Prompt | | w/Prompt | | w/PK Prompt | | Imp. Ratio (w/o→PK) | Imp. Ratio (w/→PK) |
|---|---|---|---|---|---|---|---|---|---|
| $T$ | $H$ | MSE | MAE | MSE | MAE | MSE | MAE | | |
| Dataset 1 | | | | | | | | | |
| 14 | 1 | 0.500 | 0.513 | 0.640 | 0.571 | **0.331** | **0.418** | 51.06% | 93.35% |
| | 7 | 0.806 | 0.658 | 0.774 | 0.681 | **0.588** | **0.581** | 37.07% | 31.63% |
| | 14 | 0.945 | 0.813 | 0.883 | 0.765 | **0.655** | **0.702** | 44.27% | 34.81% |
| 21 | 1 | 0.722 | 0.655 | 0.504 | 0.523 | **0.347** | **0.465** | 108.07% | 45.24% |
| | 7 | 0.753 | 0.684 | 0.824 | 0.716 | **0.563** | **0.612** | 33.75% | 46.36% |
| | 14 | 0.791 | 0.712 | 0.823 | 0.745 | **0.584** | **0.698** | 35.45% | 40.92% |
| 28 | 1 | 0.724 | 0.698 | 0.594 | 0.598 | **0.406** | **0.476** | 78.33% | 46.31% |
| | 7 | 0.728 | 0.653 | 0.838 | 0.700 | **0.491** | **0.547** | 48.27% | 70.67% |
| | 14 | 0.637 | 0.611 | 0.583 | 0.609 | **0.542** | **0.554** | 17.53% | 7.56% |
| Avg ($H = 1$) | | 0.649 | 0.622 | 0.579 | 0.564 | **0.361** | **0.453** | 79.52% | 60.33% |
| Avg ($H = 7$) | | 0.762 | 0.665 | 0.812 | 0.699 | **0.547** | **0.580** | 39.28% | 48.36% |
| Avg ($H = 14$) | | 0.791 | 0.712 | 0.763 | 0.706 | **0.594** | **0.651** | 33.24% | 28.52% |
| Dataset 2 | | | | | | | | | |
| 14 | 1 | 0.153 | 0.247 | 0.155 | 0.276 | **0.090** | **0.235** | 70.00% | 72.22% |
| | 7 | 0.203 | 0.298 | 0.206 | 0.305 | **0.155** | **0.250** | 30.97% | 32.90% |
| | 14 | 0.156 | 0.263 | 0.299 | 0.408 | **0.138** | **0.254** | 13.04% | 116.67% |
| 21 | 1 | 0.213 | 0.304 | 0.456 | 0.565 | **0.058** | **0.194** | 267.24% | 686.21% |
| | 7 | 0.223 | 0.295 | 0.203 | 0.272 | **0.169** | **0.255** | 31.95% | 20.12% |
| | 14 | 0.225 | 0.336 | 0.175 | 0.275 | **0.140** | **0.248** | 60.71% | 25.00% |
| 28 | 1 | 0.250 | 0.372 | 0.287 | 0.382 | **0.057** | **0.194** | 338.60% | 403.51% |
| | 7 | 0.370 | 0.450 | 0.290 | 0.320 | **0.191** | **0.265** | 93.72% | 51.83% |
| | 14 | 0.157 | 0.241 | 0.181 | 0.256 | **0.143** | **0.234** | 9.79% | 26.57% |
| Avg ($H = 1$) | | 0.205 | 0.308 | 0.299 | 0.408 | **0.068** | **0.208** | 200.49% | 338.05% |
| Avg ($H = 7$) | | 0.265 | 0.348 | 0.233 | 0.299 | **0.172** | **0.257** | 54.56% | 35.73% |
| Avg ($H = 14$) | | 0.179 | 0.280 | 0.218 | 0.313 | **0.140** | **0.245** | 27.79% | 55.58% |

This suggests that static or general statistical information does not provide critical forecasting value. By contrast, the third configuration, which employed the PK prompt, consistently improved performance across all input lengths and forecasting horizons. We attribute this superior performance to the inclusion of dynamically changing contextual information within the PK prompt. Whereas domain knowledge is relatively invariant and may offer limited added value across varying contexts, time-sensitive factors—such as the number of scheduled vessel arrivals, operational workload, and weather conditions—fluctuate in real-time and directly affect container terminal operations. These dynamic factors are more accurately reflected at each forecasting point through the PK prompt. Accordingly, this experiment demonstrates that incorporating dynamic contextual information via the PK prompt substantially enhances CT forecasting performance, even when an LLM-based approach is used.

In our experiments using PK prompts, we observed that the MSE improved by 79.52% and 200.49% for the two datasets when the forecasting horizon was set to one day. However, the performance gains decrease as the forecasting horizon increases. This trend can be explained by the increasing uncertainty in the contextual information provided by PK prompts over longer timeframes. For example, domain experts have noted that berth schedules typically remain stable within a 3–7-day window but are subject to frequent revisions beyond that period owing to real-time operational changes. Such volatility introduces uncertainty and reduces the reliability of contextual cues in long-term forecasts.

Nevertheless, the consistent outperformance of the model incorporating PK prompts across all experimental settings—as shown in Table IV—demonstrates the crucial role of this information in enhancing CT forecasting. If the LLM did not leverage the contextual cues embedded in the PK prompt, such consistent superiority would not have been achieved. These findings reinforce that the use of the semantic context offered by the PK prompt significantly improves the forecasting capability of LLMs in CT scenarios.

## C. Interpretation of Domain Knowledge Learning via Semantic Analysis

This section investigates two components to analyze how the LLM leverages contextual information for forecasting: (1) text prototype and (2) reprogramming layer. The two components were based on the framework proposed by [15], we selected the PK-TimeLLM model, which demonstrated the highest performance among all models for the longest input length and forecasting interval. These analyses collectively provide insights into how the LLM processes contextual information for CT forecasting from multiple perspectives.

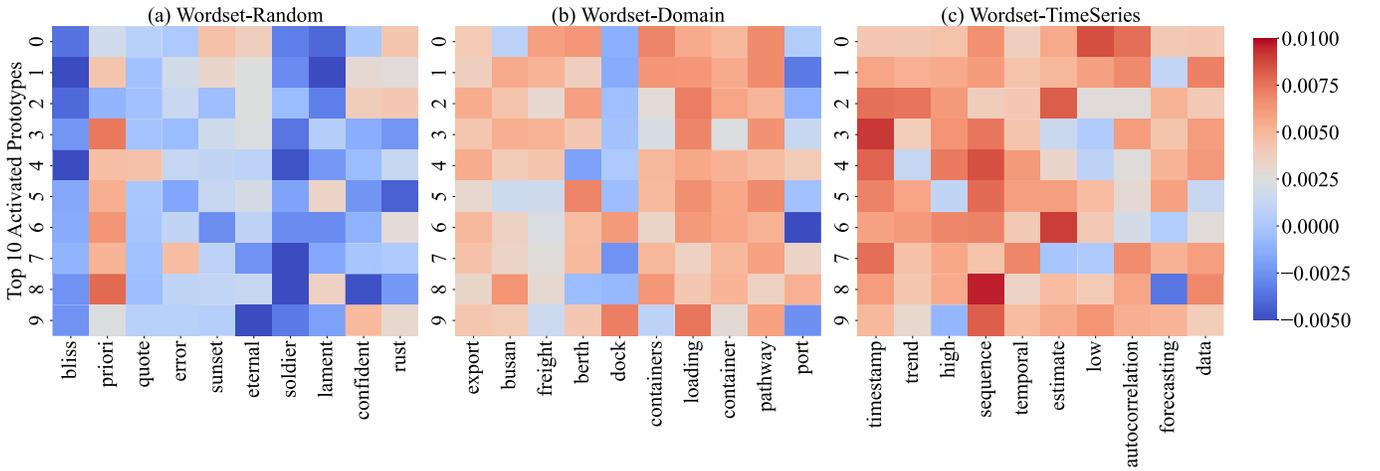

**Fig. 3.** Visualization of text prototype activation across word sets.

**1) Visualization of Text Prototypes**

Text prototype refers to the embedding vectors extracted from the word-embedding matrix and optimized for CT-forecasting tasks. If the text prototype has been effectively trained to capture contextual information, it should exhibit high activation for words semantically related to that context. To investigate this, we analyzed three word groups: (a) randomly selected words, (b) domain knowledge-related words, and (c) time-series-related (TS-related) words. Here, "activation" is defined as the sum of the weights assigned to each word when mapped into the text prototype space. We identified the top 10 prototypes with the highest activation values and visualized them. The results of this analysis are presented in Fig. 3.

Fig. 3 displays the horizontal axis as the set of words in each group, whereas the vertical axis represents the top ten text prototypes with the highest activation values for each word. Higher activation levels are shown in red, and lower activation levels are indicated in blue. Words with high activation are considered critical to the forecasting process and contribute significantly to the construction of the text prototype.

In Fig. 3, the words in the word set (a), which contained random terms, exhibited low activation overall. By contrast, word sets (b) and (c) showed high activation, confirming that random words were deemed unimportant in forming text prototypes, whereas domain knowledge-related and TS-related words were recognized as key forecasting cues. For example, words such as "berth," "container," and "loading," which showed high activation in set (b), reflect contextual knowledge indicating that CT typically increases when a vessel arrives at the "berth" and begins loading containers. This aligns with the understanding of real-world domain experts. Similarly, terms such as "timestamp," "trend," and "forecasting" in set (c) represent the temporal variability of CT and the reasoning process involved in making time-based forecasts. The purpose of the text prototype was to construct a subset of vocabulary that was highly relevant to forecasting tasks from the word embeddings. The analysis in this section demonstrates that the LLM can leverage the semantic information of forecasting-related words to build a text prototype by assigning higher weights to them. This selective weighting enables the model to extract the most relevant vocabulary.

These results suggest that the LLM can integrate contextual information—such as terminal berth operations, holidays, and CT seasonality—into the forecasting process rather than merely identifying numerical trends. This finding underscores a key advantage of LLMs over conventional TSF models: the ability to semantically interpret and incorporate diverse contextual cues that are essential for accurate forecasting.

**2) Visualization of Reprogramming Layer**

In this section, we visualize the reprogramming layer during the learning process and analyze how the alignment between text prototypes and CTs is established, as well as whether strong alignment patterns emerge. The visualization results are shown in Fig. 4.

In Fig. 4, the horizontal axis represents the text prototypes, the vertical axis denotes the CT-generated patches, and the epoch axis indicates the training progress over time. The color scale reflects the strength of alignment, with blue indicating weak alignment and red indicating strong alignment. Initially, the weights in the reprogramming layer were randomly distributed, and no discernible alignment patterns were observed. However, by epoch 4, distinct alignment patterns began to emerge, indicating that the text prototypes were being matched with relevant CT segments. This progression suggests that the LLM was actively learning which combinations of text prototypes and CT segments were the most informative for forecasting. By epoch 5, several text prototypes consistently showed strong alignment with specific CT patches, suggesting that stable and meaningful semantic associations were formed during training.

The persistent alignment suggests that the LLM consistently leverages semantic information from the same text prototypes during forecasting, indicating successful integration of contextual information with CT. If alignment with a specific text prototype is not beneficial for performance, the model would shift toward alternative alignments during training.

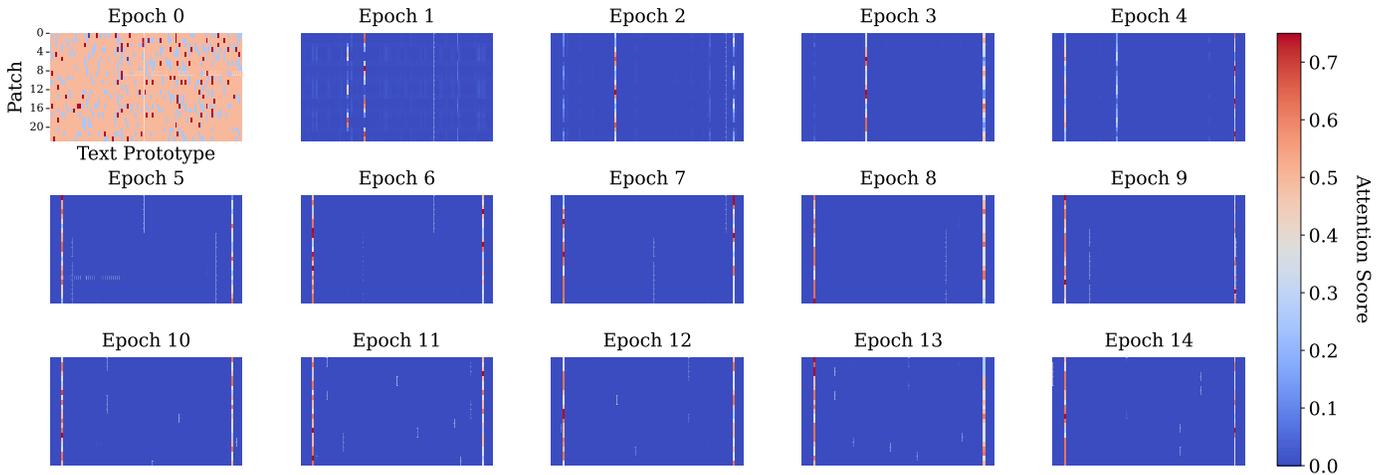

**Fig. 4.** Visualization of reprogramming layers

This analysis demonstrates that PK-TimeLLM enhances CT forecasting by effectively aligning time-series data with contextual information—a challenge that conventional TSF models struggle to address.

## V. DISCUSSION

### A. Managerial Insights

Forecasted CT functions as a key indicator for diagnosing port congestion. The periods of elevated CT tend to lengthen the gate queues of external trucks, which can precipitate gate congestion and increase the truck turnaround time (TAT). Reference [58] reported that prolonged external-truck waiting times can induce gate congestion and lead to unsatisfactory operating efficiency. To quantify the impact of CT fluctuations on TAT, we applied a simple log–log linear regression model. This specification yields elasticity coefficients that quantify the percentage change in the dependent variable associated with a 1% change in the independent variables.

We used CT as the explanatory variable and TAT as the dependent variable, where TAT was computed as the daily average dwell time of all external trucks entering the terminal. The estimates in Table V indicate that the 95% confidence interval for coefficient β is 0.261–0.474. In a log–log specification, β represents elasticity; accordingly, a 10% increase in CT is associated with an approximately 2.61–4.74% increase in TAT.

TABLE V
REGRESSION RESULTS OF CT ON TAT

| Predictor | β | 95% CI [Lower, Upper] | z | p-value |
|---|---|---|---|---|
| Intercept | 3.476 * | [-0.064, 7.016] | 1.925 | < .05 |
| ln(CT) | 0.367 *** | [0.261, 0.474] | 6.757 | < .001 |

Because TAT is incurred per vehicle, increases in TAT accumulate across all active trucks. For delivery companies managing many trucks, this accumulation leads to large productivity losses. Consequently, delivery companies can leverage CT forecasts when scheduling inbound and outbound container movements. If an elevated CT is anticipated at a particular terminal, gate-ins can be steered toward days or windows with lower CT at that terminal while reprioritizing assignments toward other terminals to increase inland transportation productivity. In terminal operations and planning, the forecasted CT serves as a key input for determining capacity caps in the gate appointment system. Reference [58] argued that if the future workload can be forecasted, the time-of-day booking limits in a truck appointment system can be rationally determined. Accordingly, combining the forecasted CT with the terminal's service capacity enables the dynamic allocation of appointment slots and resource scheduling so that congestion remains within acceptable bounds while operational efficiency is maximized. In particular, when a surge in CT is anticipated and residual service capacity is tight, appointment caps may be lowered; conversely, when slack is expected, caps may be raised to promote temporal and inter-terminal workload smoothing, thereby mitigating congestion in port operations.

### B. Limitations

This study did not conduct a comprehensive comparative analysis to evaluate the effectiveness of the proposed PK prompt and selected LLM architecture. Although Section IV.B.3 demonstrated that PK prompts incorporating dynamic contextual information significantly enhanced CT-forecasting performance compared with prompts using static data, we did not quantitatively assess whether variations in prompt structure—such as the sequence or content of textual components—could further improve outcomes. Similarly, although we adopted the relatively lightweight Qwen 3B model, it remains to be determined whether the model performance follows a scaling law or whether a simpler structure would suffice. Future investigations should include quantitative comparative studies across various LLM architectures and prompt design configurations to improve the understanding of the generalizability and scalability of LLM-based forecasting frameworks.

| Model | Category | Parameters | | | Training Time (s/iteration) | Number of Inferences per Second |
|---|---|---|---|---|---|---|
| | | Trainable | Non-Trainable | Total | | |
| LSTM | TSF | 12.05 MB | - | 12.05 MB | 0.185 | 41.66 |
| LSTnet | TSF | 1.53 MB | - | 1.53 MB | 0.272 | 40.00 |
| DLinear | TSF | 0.00 MB | - | 0.00 MB | 0.136 | 100.00 |
| PatchMixer | TSF | 0.12 MB | - | 0.12 MB | 0.361 | 41.66 |
| PatchTST | TSF | 0.18 MB | - | 0.18 MB | 0.185 | 40.00 |
| TSMixer | TSF | 0.01 MB | - | 0.01 MB | 0.193 | 55.55 |
| AutoTimes | LLM-TSF | 474.71 MB | - | 474.71 MB | 2.675 | 5.05 |
| aLLM4TS | LLM-TSF | 324.48 MB | 150.24 MB | 474.72 MB | 2.323 | 3.74 |
| PK-TimeLLM | LLM-TSF | 82.20 MB | 11,771.92 MB | 11,854.12 MB | 21.073 | 0.32 |

Another critical limitation of the adoption of LLMs is their substantial demand for computational resources during both training and inference. This constraint is highlighted in Table VI, which reports the computational efficiency of the models evaluated in this study. The table presents the number of trainable and non-trainable parameters, along with the training and inference durations, based on experiments conducted using two L40S GPUs, each equipped with 48 GB of VRAM.

The results clearly demonstrate that LLM-TSF models require significantly more parameters than conventional TSF models. For example, the parameter count of the nonprompted LLM-TSF model was approximately 39.5 times greater than that of the LSTM model, which had the highest parameter count among the conventional models. Notably, the PK-TimeLLM model contains nearly 988 times more parameters than the LSTM model, underscoring the immense resource requirements of LLM-based forecasting methods.

These findings suggest that despite the superior predictive performance of LLM-based approaches, the trade-off between accuracy and computational cost raises concerns regarding their practical and economic viability. Notably, the PK-TimeLLM model employed in this study forecasts daily CT, implying that inference is performed only once per day. Therefore, a longer inference time may still be acceptable when accompanied by substantial performance gains. However, within the context of port logistics, numerous operational tasks—such as optimization and scheduling, yard allocation, vessel arrival and turnaround prediction—must be managed simultaneously. Some of these tasks require predictions at the container level and may need to be executed multiple times per second, making the deployment of LLM-based inference in real-world operations prohibitively difficult. From a practitioner's standpoint, it may be more feasible to utilize lightweight models that offer moderate accuracy but require fewer computational resources rather than deploying state-of-the-art LLMs with high resource demands. This limitation constitutes a substantial barrier to the widespread adoption of LLMs in port operations.

### C. Future works

Building on the findings and limitations of this study, several directions for future research can be identified. First, advancing the data infrastructure for LLM-based port logistics research is essential. This includes constructing a domain-specific corpus that converts operational records—typically stored in relational databases—into hierarchical, natural language representations spanning container-level events, daily operational summaries, and equipment productivity indicators. Such a corpus would enable LLMs to perform sentence-level reasoning over port activities. In addition, integrating fragmented data sources exchanged among multiple stakeholders—such as Automatic Identification System data, Electronic Data Interchange records, and customs clearance statuses—would allow LLMs to comprehensively understand container-level logistics flows. Furthermore, LLM-based data standardization and harmonization offer a promising approach to normalizing the heterogeneous and unstandardized data prevalent in port operations, thereby improving overall data quality and enabling effective cross-port data integration.

Second, optimizing the trade-off between model scale and computational efficiency is critical for the practical deployment of LLMs in port logistics. Most ports lack the infrastructure to support large-scale LLMs, and container-level operations often demand real-time responsiveness. LLM compression techniques have shown potential to substantially reduce computational requirements; however, identifying the optimal balance between model complexity and task performance remains an open challenge. Future research needs to conduct quantitative evaluations of forecasting performance as a function of model size and develop lightweight architectures that maintain acceptable accuracy under the operational constraints of real-world port environments.

### VI. CONCLUSION

This study proposed an LLM-based approach to integrate diverse types of contextual information into CT forecasting—an area still underexplored within port logistics. A novel prompt structure, referred to as the PK prompt, was introduced to guide the LLM in capturing domain-specific knowledge. Through comparative experiments using real operational data and a suite of benchmark models, we demonstrated that the proposed method achieves state-of-the-art performance compared with recent TSF approaches. Furthermore, our findings indicate that the LLM effectively learns and leverages various forms of contextual information that are unique to port-logistics operations. In addition, we examined limitations and outlined future research directions, offering new insights into the intersection of LLMs and port logistics. These contributions directly address the research questions introduced at the beginning. First, the LLM-based approach outperformed conventional TSF models in CT forecasting. Second, the results confirmed that LLMs can learn and interpret complex contextual information that is specific to port operations. Third, despite these promising results, several challenges remain in the practical implementation of LLMs in operational settings. Continued research efforts are essential to bridge the gap between theoretical advancements and real-world implementation.

Generative AI, including LLMs, is driving innovation across diverse sectors, including port logistics, by introducing novel, data-driven solutions. This study provides a foundational perspective on the application of LLMs in port logistics by presenting real-world use cases and outlining future research directions. This initiative is expected to support the development of domain expertise in forecasting, operational optimization, and decision automation. This may contribute to enhancing the productivity and sustainability of port-logistics systems.

# APPENDIX

## APPENDIX A
### SEARCH SPACE OF HYPERPARAMETERS

| Model | Architecture | Hyper parameter | Search space |
|---|---|---|---|
| LSTM | Recurrent | num_layers<br>hidden_size | [1 2]<br>[32 64 128 256] |
| LSTnet | Hybrid | hidden_CNN<br>hidden_RNN<br>kernel_size<br>highway window<br>skip<br>hidskip | [128 256]<br>[128 256]<br>[3 6]<br>[0 24]<br>[3 6]<br>[3 6] |
| PatchMixer | Convolution | num_layers<br>n_heads<br>d_model<br>d_ff | [1 2 3]<br>[4 8]<br>[32 64]<br>[128 256] |
| PatchTST | Transformer | num_layers<br>kernel_size<br>n_heads<br>d_model<br>d_ff | [1 2 3]<br>[1 3 5]<br>[4 8]<br>[32 64]<br>[128 256] |
| DLinear | Linear | kernel_size | [1 3 5] |
| TSMixer | MLP | num_layers | [1 3 5] |
| AutoTimes | LLM | mlp_hidden_dim<br>mlp_hidden_layers | [128 256]<br>[2 4] |
| aLLM4TS | LLM | llm_layers<br>n_heads<br>d_ff | [2 4]<br>[4 8]<br>[128 256] |
| TimeLLM | LLM | num_tokens<br>n_head<br>d_model<br>d_ff | [100 1,000]<br>[4 8]<br>[32 64]<br>[128 256] |

## APPENDIX B
### EXAMPLE OF W/ PROMPT

| Category | Contents |
|---|---|
| Domain | This dataset captures the daily CT, measured by gate-in and gate-out activities at a container terminal in Busan Port, spanning the period from January 2022 to December 2022. |
| Instruction | Forecast the next $< H >$ steps given the previous $< T >$ steps information attached. |
| Domain Information | Berth schedule data are used to identify the estimated time of arrival of vessels and the anticipated CT. CT generally increases with the number of scheduled vessel arrivals, as a higher number of vessels typically requires more operational activity. Throughput often declines on weekends and public holidays owing to reduced road truck activity and may also decrease on days with heavy rainfall due to weather-related disruptions. |
| Statistics | min value $< min(X_t) >$ , max value $< max(X_t) >$, median value $< median(X_t) >$, the trend of input is $[upward|downward]$, top 5 lags are :$< lags >$ |